\documentclass[a4paper,12pt]{article} 
\usepackage{graphicx} 
\usepackage{graphics} 
\usepackage{amsmath} %needed for \begin{align}... \end{align} environment
\usepackage{amsfonts}  %necessaire aux caracteres spéciaux
\usepackage{amssymb} %necessaire aux caracteres spéciaux 

\usepackage[colorlinks=true, pdfstartview=FitV, linkcolor=blue,  citecolor=blue, urlcolor=blue]{hyperref}  %liens hypertexte

\newcommand{\Cs}{\mathbb{C}} %C scale
\newcommand{\Us}{\mathbb{U}} 
\newcommand{\Vs}{\mathbb{V}} 
\newcommand{\Ws}{\mathbb{W}} 
\newcommand{\Pl}{\mathbb{P}} %Planck

\def\l{\left}
\def\r{\right}
\def\beq{\begin{equation}}
\def\eeq{\end{equation}}
\def\d{\partial}

\begin{document}
%%%%%%%%

\title{Scale-relativistic correction to the muon anomalous magnetic moment}
%titre provisoire

\author{Laurent Nottale\\
{\small \it LUTH, Observatoire de Paris-Meudon, F-92195 Meudon, France}\\
{\small laurent.nottale@obspm.fr}}
\maketitle

\begin{abstract}
%%%%%%%%
The anomalous magnetic moment of the muon is one of the most precisely measured quantities in physics. 
%v3
Its experimental value exhibits a $4.2 \, \sigma$ discrepancy $\delta a_\mu=(251 \pm 59) \times 10^{-11}$ with its theoretical value calculated in the standard model framework, while they agree for the electron. The muon theoretical calculation involves a mass-dependent contribution which comes from two-loop vacuum polarization insertions due to electron-positron pairs and depends on the electron to muon mass ratio $x=m_e/m_\mu$. In standard quantum mechanics, mass ratios and inverse Compton length ratios are identical. This is no longer the case in the special scale-relativity framework,
 in which the Planck length-scale is invariant under dilations. %v3
Using the renormalization group approach, we differentiate between the origin of $ \ln x$ logarithmic contributions which depend on mass, and $x$ linear contributions which we assume to actually depend on inverse Compton lengths. 
%v3
By defining the muon constant $\Cs_\mu=\ln(m_\Pl/m_\mu)$ in terms of the Planck mass $m_\Pl$, the resulting scale-relativistic correction writes $\delta a_\mu= -\alpha^2 \, (x \:\ln^3 x)/(8 \; \Cs_\mu^2)$, where $\alpha$ is the fine structure constant. Its numerical value, $(230 \pm 16)  \times 10^{-11}$, is in excellent agreement with the observed theory-experiment difference. 
\end{abstract}
 %%%%%%%

The Dirac equation predicts a muon magnetic moment, $M_\mu= g_\mu (e/2 m_\mu) S$, with
gyromagnetic ratio $g_\mu= 2$. Quantum loop effects lead to a small calculable deviation, parameterized by the anomalous magnetic moment (AMM) $a_\mu=(g_\mu-2)/2 $ \cite{PDG2020}.
 
That quantity can be accurately measured and precisely predicted within the framework of the Standard Model (SM). Hence, comparison of experiment and theory tests the SM at its quantum loop level. 

While the theoretical and experimental values of the electron anomalous magnetic moment agree within uncertainties \cite{Mohr2016, Kinoshita2012},
\beq
a_e({\rm th})=0.00115965218178(77), \;\;\; a_e({\rm exp})=0.00115965218091(26),
\eeq
on the contrary, the muon anomalous magnetic moment exhibits, since about 20 years \cite{PDG2020,Jegerlehner2009,Jegerlehner2018}, a difference of $(268 \pm 72) \times 10^{-11}$, statistically significant at the $3.7 \, \sigma$ level:
\beq
a_\mu({\rm th})=116591823(36) \times 10^{-11}, \;\;\; a_\mu({\rm exp})=116592091(63) \times 10^{-11}.
\eeq
This effect constitutes one of the main discrepancies between the SM theoretical predictions and experiments. It is all the more puzzling that it is not found in the high energy realm of today's particle physics frontier ($\approx 10$ TeV), but instead at the atomic-nuclear scale ($\approx 100$ MeV) which was up to now thought to be fully understood.

%v3 et extended
More recently, the first results of the Fermilab National Accelerator Laboratory (FNAL) Muon $g-2$ E989 Experiment \cite{Abi2021} have been found to be in excellent agreement with the previous BNL E821 measurement. They obtained $a_\mu({\rm FNAL})=116592040(54) \times 10^{-11}$, leading to a combined BNL+FNAL average result:
\beq
a_\mu({\rm exp})=116592061(41) \times 10^{-11},
\eeq
increasing the tension between experiment \cite{Abi2021} and theory \cite{Aoyama2020}, $\delta a_\mu=(251 \pm 59) \times 10^{-11}$, to 4.2 standard deviations.
%fin v3
 
We suggest here that the muon magnetic moment discrepancy could be explained in the special scale-relativity (SSR) framework by accounting for the correction to the relation between mass ratios and Compton length ratios which occurs in this theory.
 
Recall that the principle of relativity, which was up to now applied, in current theories of relativity, to origin, orientation and motion transformations of the coordinate system, has been extended to apply also to scale transformations of the measurement resolution, which is added to the variables that characterize its (relative) state \cite{Nottale1989,Nottale1992,Nottale1993,Nottale2011}.
 
  The scale-relativity theory is the general framework built from this first principle \cite{Nottale1993,Nottale2011}, including the construction of new scale laws of log-Lorentzian form \cite{Nottale1992}, a geometric foundation of quantum mechanics \cite{Nottale2007} and of gauge fields \cite{Nottale2006} in terms of a nondifferentiable and continuous fractal space-time
whose geodesics define wave functions, %v3
and the suggestion of the possibility of a new macroscopic quantum-type mechanics (based on a constant different from $\hbar$ and relevant to chaotic systems beyond their horizon of predictability and to turbulence) \cite{Nottale1993,Nottale1996,Nottale1997,Auffray2008,Nottale2008,Turner2015,Nottale2018}.

 Here we are concerned with only the log-Lorentzian scale laws aspect of the theory, specifically, with the fact that inverse length-scales and mass-scales, which are identical in standard quantum mechanics (QM), become different in the SSR framework.
 
 We have mathematically proved \cite{Nottale1992}, \cite[Chapt.~6]{Nottale1993}, \cite[Chapt.~4.4]{Nottale2011} that the general solution to the special relativity problem (i.e., find the linear laws of transformations which come under the principle of relativity) is, as well for motion as for scales, the Lorentz transformation. This proof is based on only two axioms, internality of the composition law and reflection invariance, which are both expressions of the only principle of relativity. We know since Poincar\'e and Einstein that the special relativity law of composition of two velocities $u$ and $v$ writes $w={(u+v)}/{(1+u \, v / c^2)}$.
 In the same way, the general law of composition of length-scale ratios, $r \to r'= \rho \times r$, writes in special scale-relativity theory
$\Ws={(\Us+\Vs)}/{(1+\Us \Vs/\Cs^2)}$,
where $\Us=\ln(r/\lambda)$, $\Vs=\ln \rho$ and $\Ws=\ln(r'/\lambda)$ 
%v3
and $\Cs$ is a constant whose meaning will be specified in the following. 

 Another relevant result of special motion-relativity is that the Galilean relation of equality between velocity and momentum, $p/mc=v/c$, becomes
\beq
\frac{p}{mc}=\frac{v/c}{\sqrt{1-(v/c)^2}}.
\eeq
In the same way as velocity characterizes the state of motion of the reference system, we consider in SSR that resolution length-scales characterize their state of scale \cite{Nottale1989,Nottale1992,Nottale1993,Nottale2011}. The difference is that motion transformations constitute an additive group, while scale transformations are a multiplicative group. However, using the Gell-Mann-Levy method one can show \cite{Nottale1993,Nottale2011} that the natural variable for describing length-scales and their transformations is the logarithm of a ratio of length intervals, so that one recovers an additive group in terms of these variables.

In analogy with the various levels of theories of motion-relativity, one can first define a ``Galilean scale-relativity" (GSR) framework, which just corresponds to the usual laws of dilation and contraction. The standard de Broglie law which relates momentum and length-scales in quantum mechanics, $p=\hbar/\lambda$, can be established from Noether's theorem in this framework. This is similar to the obtention,  in motion-relativity theory, of the standard relation $p=mv$ from uniformity of space.  It may be also written as a direct equality $\ln(p/p_0)=\ln(\lambda_0/\lambda)$, with $p_0 \lambda_0=\hbar$. The generalization of this relation in SSR involves a log-Lorentz factor \cite{Nottale1992,Nottale1993,Nottale2011}:
\beq
 \ln \frac{p}{p_0} = \frac{     \ln({\lambda_0}/{\lambda})     } 
 {    \sqrt{   1-  \ln({\lambda_0}/{\lambda})^2   /   \Cs^2   }    }.
\eeq
The usual GSR law is clearly recovered in the limit $\Cs \to \infty$. The meaning of this constant can be clarified by expressing it also in terms of the reference scale $\lambda_0$:
\beq
\Cs_{0}=\ln \frac{\lambda_0}{\lambda_\Pl}.
\eeq
 This introduces a minimal scale $\lambda_\Pl$ which is invariant under dilations and contractions, unreachable and unpassable, whatever the scale $\lambda_0$ which has been taken as reference \cite{Nottale1992,Nottale1993}. Moreover, momentum-energy now tends to infinity when the length-time scale tends to this limit, which therefore plays the role of the zero scale interval of the standard theory. This remarkable property has naturally led us \cite{Nottale1992} to identify it with the Planck length-scale,
 \beq
 \lambda_\Pl=\sqrt{  \frac{\hbar \: G}{c^3}  }.
 \eeq
 
 Let us now analyse the current situation of the muon AMM in the light of the SSR framework. The fact that there is no strong effect on the electron but only on the muon points toward a manifestation of the mass-scale increase by a factor $\approx 200$ between the electron and the muon, which is the only difference between these two particles (apart from their consequently different lifetimes). 
 
 %v3: on distingue maintenant échelle de transition et échelle de référence
In order to apply the SSR framework to the muon AMM problem, we need to specify the nature of two fundamental scales. The first is the scale at which occurs the transition between Galilean scale relativity (GSR) and Lorentzian special scale relativity (SSR).  
%fin v3
A natural identification of this transition is with the Compton scale of the electron $\lambda_e$ \cite{Nottale1992,Nottale1993,Nottale2011}. Indeed, physics changes drastically at scales smaller than $\lambda_e$, in a way that is directly related to our purpose: namely, the various physical quantities, in particular masses and charges, become explicitly dependent on scale below $\lambda_e$, a behaviour that is currently accounted for in terms of vacuum polarization and radiative corrections and well described by the renormalization group equations.

%v3
The second fundamental scale that should be known is the reference scale defining the constant $\Cs_0$. It is naturally given by the Compton length of the particle studied \cite{Nottale1992}. 
When the particle considered is the electron, both scales coincide. But here we are concerned with the muon, so that the SSR constant is given by the Compton length of the muon, i.e. $\Cs_\mu=\ln(\lambda_\mu / \lambda_\Pl) \approx \ln (m_\Pl/ m_\mu)=46.196$. This is a new configuration, in which the electron Compton length is larger than the reference length. This is justified by the fact that the $e^+e^-$ pairs considered here are virtual, i.e. they are interpreted as being part of the set of fractals geodesics which constitutes the muon as a whole. 
%finv3
 
 A possible intervention of SSR corrections can therefore be expected, since the ratio of the muon and electron Compton lengths is slightly different from their inverse mass ratio in this framework. Taking the muon scale as reference scale, it is given by:
 %v3=C_e-> C_mu
 \beq
 \ln \frac{\lambda_\mu}{\lambda_e} = \frac{\ln({m_e}/{m_\mu})} 
 {\sqrt{1+\ln^2({m_e}/{m_\mu})/\Cs_\mu^2}}.
 \label{eq8}
 \eeq
Therefore the length-scale ratio becomes $y=\lambda_\mu /\lambda_e= x^{[1+(\ln x/\Cs_\mu)^2]^{-1/2}}\approx 1/200$ instead of the mass ratio $x={m_e}/{m_\mu}\approx 1/207$. 
 
 %1%%%%%%%
\begin{figure}[!ht]
\begin{center}
\includegraphics[width=5cm]{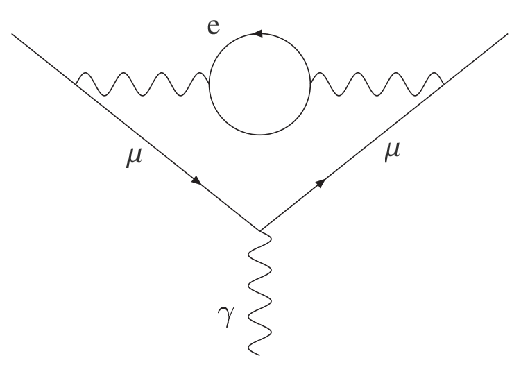}
\caption{\small{Feynman diagram for the leading mass-dependent term,
%v3
involving $e^+ e^-$ pairs,  of the fourth-order radiative correction to the scattering of a muon by an electromagnetic field, which plays a central role in the SSR correction to the theoretical calculation of the muon AMM.}}
\label{fig1}
\end{center}
\end{figure}
%%%%%%%%

 As a direct consequence, one is led to look for a possible SSR effect only among the mass-dependent terms of the muon AMM calculation. There is no such contribution to 1-loop level, i.e. in Schwinger's original $\alpha/2 \pi$ correction to the magnetic moment  \cite{Jegerlehner2009}. To 2-loop, the main mass-dependent and electron-dependent  term comes from the Feynman diagram shown in Fig.~\ref{fig1}, which corresponds to the insertion of a closed lepton loop (electron-positron pair).
 %Jegerlehner p. 25 et suiv

Let us first recall how the standard calculation of this contribution is performed without SSR correction. It has been calculated in 1949 by Karplus and Kroll (KK) \cite{Karplus1950}. It reads in position space (we denote by $X$ the position coordinate so as not to confuse it with $x=m_e/m_\mu$ in what follows):
\beq
 \bar{M}^{IIe}=-\frac{e \alpha^2}{4 \hbar c} \int d^4X_0 d^4X_1 d^4X_2 A_\mu^e(X_0) \bar{D}_F^{(2)}(X_2-X_1) \bar{\psi}(X_1) \gamma_\nu S_F(X_0-X_1) \gamma_\mu S_F(X_2-X_0) \gamma_\nu \psi(X_2).
\eeq
The bar on $\bar{M}^{IIe}$ and $\bar{D}_F$ indicates that the renormalization terms have been removed. The Feynman function ${D}_F$ describes the properties of a virtual photon as modified by its interaction with the electron-positron field. The leading term is obviously
\beq
D_F^{(1)}(X)=-\frac{2i}{(2\pi)^4}\int e^{-ipX} \; \frac{d^4p}{p^2}.
\eeq
The radiative corrections to this function arise from the ability of the virtual photon to create pairs.  The first term $D_F^{(2)}$ is simply due to the creation and annihilation of one pair, as can be seen in Fig.~\ref{fig1}. After removal of the renormalization term and performing its Fourier transform, it reads
\beq
 \bar{D}_F^{(2)}(p)=\frac{\alpha}{2 \pi} \int_0^1 dv \; \frac{2 v^2(1-v^2/3)}{p^2(1-v^2)+4 \kappa_e^2},
\eeq
where $\kappa_e$ is the electron mass term. Feynman's $S_F(X)$ function describes the propagation of the particle for which the magnetic moment is computed (by looking at its elastic scattering by an electromagnetic field). This particle is an electron in the original KK calculation and a muon in the case considered here:
 \beq
 S_F(X,\kappa_\mu)=-\frac{2i}{(2 \pi)^4} \int d^4 p \;e^{-i\,p\,X} \; \frac{i \, \gamma \,  p- \kappa_\mu}{p^2+\kappa_\mu^2},
 \label{SF}
 \eeq
where the mass term $\kappa_\mu$ is now the muon one.

The subsequent calculations are continued by KK in momentum space. The momentum $p_1$ is used to denote the momentum of the final state and $p_2$ the momentum of the initial state:
\begin{eqnarray}
\psi(X)=\int e^{i p_2 X} \psi(p_2) d^4 p_2, \;\;\; (i \gamma p_2+ \kappa_\mu) \psi(p_2)=0,\\
\bar{\psi}(X)=\int e^{i p_1 X} \psi(p_1) d^4 p_1, \;\;\; \bar{ \psi}(p_1)(i \gamma p_1+ \kappa_\mu)=0.
\end{eqnarray}
Then KK find, for the contribution considered, the final expression:
\begin{eqnarray}
 \bar{M}^{IIe}=-\frac{2 i e \alpha^2}{ \hbar c} \int d^4 p_1 d^4 p_2  A_\mu^e(p_1-p_2)
  \int d^4 k \int_0^1 dv 
  \frac{2 v^2 (1-v^2/3)}{k^2(1-v^2)+4 \kappa_e^2} \nonumber\\
 \times \bar{\psi} (p_1)\gamma_\nu
  \frac{i \, \gamma \,  (p_1-k)- \kappa_\mu}{(p_1-k)^2+\kappa_\mu^2}
  \gamma_\mu
   \frac{i \, \gamma \,  (p_2-k)- \kappa_\mu}{ (p_2-k)^2-\kappa_\mu^2}
   \gamma_\nu \psi(p_2),
\end{eqnarray}
in which we have now clearly identified the muon mass term $\kappa_\mu$ and the electron-positron pair mass term $2 \kappa_e$.

 The resulting muon vacuum polarization insertion was explicitly computed in the late 1950's \cite{Suura1957,Petermann1957}. It is given by the double integral:
 \beq
 a_\mu^{(4)}=\l(\frac{\alpha}{\pi}\r)^2\int_0^1 du \; u^2 (1-u)  \int_0^1 dv \; \frac{v^2 (1-v^2/3)}{u^2(1-v^2)+4 x^2 (1-u)},
 \label{Suura}
 \eeq
 where $x=\kappa_e/\kappa_\mu=m_e/m_\mu$ is the electron to muon mass ratio
 in the SM framework. %v3
An exact integration has been performed by Elend in 1966 \cite{Elend1966} and the final expression has been written in compact form by Passera \cite{Passera2007} as:
 \begin{align}
 \begin{split}
a_\mu^{(4)}=\l(\frac{\alpha}{\pi}\r)^2
 \l( 
 -\frac{25}{36}   -\frac{1}{3}\ln x
+x^2 (4+3 \ln x)     
+x^4 \left[    \frac{\pi ^2}{3}-2  \ln x\ln \left(\frac{1}{x}-x\right)-\text{Li}_2\left(x^2\right)     \right]    \right.  \\  \left. 
+ \frac{x}{2} \left(1-5 x^2\right)   \left[ \frac{\pi ^2}{2}-\ln x \ln \left(\frac{1-x}{1+x}\right)-\text{Li}_2(x)+\text{Li}_2(-x) \right]  
   \r).
 \end{split}
 \label{eq9}
 \end{align}
 A complete expansion of this expression has been given by Li et al. \cite{Li1993}, for both cases $x>1$ and $x<1$. Keeping only the leading terms, it yields in terms of $\ln x$ and $x$:
 \beq
 a_\mu^{(4)}=\l( \frac{\alpha}{\pi} \r)^2  \l(  -\frac{25}{36}   -\frac{1}{3} \ln x + \frac{\pi^2}{4} x  +{\cal O}(x^2 \ln x) \r).
 \label{eq10}
 \eeq
 
 In the SSR framework, this formula can no longer be correct. Indeed, the muon Compton length-scale and mass scale, when they are referenced to the electron scale, are no longer strictly inverse quantities. One should therefore make the difference between $x=m_e/m_\mu$ and $y=\lambda_\mu/\lambda_e$, whose relationship is given in Eq.~(\ref{eq8}).
 
 An analysis of the way the above mass dependent contribution to the muon AMM is obtained shows that it depends on {\it both} mass and Compton length. The original KK work starts from position space then ends the calculation in momentum space through the standard assumption of Fourier transform between position and momentum representations. This calculation relies on the previous work of Dyson \cite{Dyson1949}, who explicitly specifies that the mass term $\kappa$ in Feynman's $S_F(X)$ function (Eq.~\ref{SF}) is given by the reciprocal Compton length, not the mass itself. On the other hand, we also know that masses enter as such in the AMM calculation, in particular through the threshold $2m_e$ for pair creation.

However, the two contributions are not separated in the KK calculation of the electron AMM, nor in the muon calculation of Suura-Wichmann \cite{Suura1957} and Peterman \cite{Petermann1957} derived from it. We therefore need to use another approach where the two contributions are separated. The renormalization group approach provides us with such a separation. It has been shown by Lautrup and de Raphael  \cite{Lautrup1974} that the equation for the contribution to the muon AMM from electron vacuum polarization insertions takes the form of a Callan-Symanzik equation
\beq
\l( m_e \frac{\d}{\d m_e} + \beta(\alpha) \frac{\d}{\d \alpha} \r) a_\mu \l(\frac{m_e}{m_\mu},\alpha \r) =R\l(\frac{m_e}{m_\mu}\r),
\label{eq11}
\eeq
where the right hand side of this equation is found to be vanishing as $x=m_e/m_\mu$ instead of the naive expectation $x^2$ from Eq.~(\ref{Suura}). In this expression, no difference is yet made between the mass ratio and inverse Compton length ratio. 
However, the leading logarithmic term of Eq.~(\ref{eq10}),
$-\frac{1}{3} \ln x$, %v3
 is now provided by the left hand side of this equation \cite{Jegerlehner2009}, while the linear term
 $\frac{\pi^2}{4}\, x$  %v3
comes from its right hand side.

In the l.h.s. of Eq.~(\ref{eq11}), $\beta(\alpha)$ is the standard QED $\beta$-function, $\beta(\alpha)=2 \alpha^2/3 \pi$ to one-loop. The meaning of this equation is that the origin of the leading logarithm term $-\frac{1}{3}\ln x$ is just charge screening of the electromagnetic charge \cite{Jegerlehner2009} and that it comes from the mere running QED coupling at muon scale,
\beq
\alpha_\mu=\alpha \l(1+ \frac{2 \alpha}{3\pi} \ln\frac{m_\mu}{m_e}\r).
\eeq
This contribution is therefore generated by the electron-muon mass ratio $x={m_e}/{m_\mu}$. 

On the contrary, if one now considers the r.h.s. of Eq.~(\ref{eq11}), a detailed analysis by Lautrup and de Raphael \cite{Lautrup1974} shows that it finds its origin in the $S_F(X)$ functions in which the mass terms are actually defined as inverse Compton lengths \cite{Dyson1949}. 
We are therefore led to differentiate between $ \ln x$ logarithmic contributions which depend on mass, and $x$ linear contributions which we assume to actually depend on inverse Compton lengths.%v3

Therefore the Callan-Symanzik equation for the muon AMM can now be written (keeping only the leading terms) as
\beq
\l( x \frac{\d}{\d x} + \beta(\alpha) \frac{\d}{\d \alpha} \r) a_\mu(x,\alpha) =\l(\frac{\alpha}{2} \r)^2 y,
\eeq
where $y=\lambda_\mu/\lambda_e$ while $x=m_e/m_\mu$.

%v3
 The solution to order $\alpha^2$ of this equation finally yields a correction to the muon AMM, $\delta a_\mu^{\rm SSR}=(y-x)\alpha^2/4$, which is well approximated by
 \beq
 \delta a_\mu^{\rm SSR}= - \alpha^2 \;  \frac{x \:\ln^3 x}{8 \: \Cs_\mu^2}  =230 \times 10^{-11} .
 \eeq
Assuming that higher order power contributions $x^k$ should also be corrected and replaced by $y^k$, one finds that %v3
the 2-loop $x^2$ power terms of Eq.~(\ref{eq9}), coming from $e^+e^-$ pairs, yield a small correction $\delta a_\mu=4 \times 10^{-11}$, while $x^2 \ln x$ terms yield $-20  \times 10^{-11}$. Hadron loops contribute also by $\sim (m_\mu/m)^2$ terms, yielding a SSR correction $\approx +17 \times 10^{-11}$, so that these possible higher order effects cancel each other and are anyway smaller than current theoretical uncertainties ($43 \times 10^{-11}$). Three-loop effects as well as tau lepton loops yield corrections smaller than $1 \times 10^{-11}$.  

We therefore estimate the special scale-relativistic correction to $\delta a_\mu=(230 \pm 16)  \times 10^{-11}$, yielding a theoretically estimated  (SM+SSR) muon AMM
\beq
a_\mu({\rm th})=116592040(43) \times 10^{-11},
\label{newvalue}
\eeq
in excellent agreement with the present experimental value $a_\mu({\rm exp})=116592061(41) \times 10^{-11}$ \cite{Abi2021}). Provided the SM theoretical result is confirmed in the future (note however that some QCD lattice calculations disagree with it and yield a higher SM theoretical prediction \cite{Borsanyi2021}, although with a larger error bar), the SSR correction $(230 \pm 16)  \times 10^{-11}$ (where the error bar corresponds here to the possible inclusion of higher order mass terms) agrees with the observed difference between the experimental and theoretical muon AMM,  $\delta a_\mu=(251 \pm 59)  \times 10^{-11}$.
 
 Moreover, it will be possible to test the newly estimated theoretical value of the muon AMM Eq.~(\ref{newvalue}) in the next years, since the Muon $g-2$ Fermilab experiment E989 aims to finally reduce the experimental uncertainty by a factor of four \cite{Grange2015}. If this goal is attained (experimental error $\approx 13 \times 10^{-11}$), the main uncertainty in the comparison between theory and experiment will come from the hadronic theoretical contribution, for which improvements are also expected \cite{Keshavarzi2020}.
 
 We intend to perform in a forthcoming work a complete new theoretical calculation of the muon AMM in the SSR framework, which involves taking into account a generalized form of Fourier transform \cite[p.~540]{Nottale2011} compatible with the new status of the Planck length as a minimal scale, invariant under dilations.

%%%%%%%%%%%%
%{\bf Acknowledgements}: 
%%%%%%%%%%%%

%%%%%%%%%%%%

%%%%%%%%%%%
   
%%%%%%%%

\begin{thebibliography}{}

\bibitem{PDG2020} P.A. Zyla et al. (Particle Data Group), {\it Prog. Theor. Exp. Phys.} 2020, 083C01 (2020).
%
\bibitem{Mohr2016} P.J. Mohr, D.B. Newell and B.N. Taylor, (CODATA 2014), {\it Rev. Mod. Phys.} {\bf 88}, 1 (2016).
 %
 \bibitem{Kinoshita2012} T. Aoyama, M. Hayakawa, T. Kinoshita, and M. Nio, {\it Phys. Rev. Lett.} {\bf 109}, 111807 (2012), arXiv:1205.5368.
%
\bibitem{Jegerlehner2009} F. Jegerlehner and A. Nyffeler, {\it Phys. Rep.}, {\bf 477}, 1 (2009).
%
\bibitem{Jegerlehner2018} F. Jegerlehner, {\it Acta Phys. Pol.} {\bf 49}, 1157 (2018).
%
\bibitem{Abi2021} B. Abi et al., {\it Phys. Rev. Lett.} {\bf 126}, 141801 (2021).
%
\bibitem{Aoyama2020} T. Aoyama et al. [Muon g-2 Theory Initiative], {\it Phys. Rept.} {\bf 887}, 1-166 (2020). 
%
\bibitem{Nottale1989} L. Nottale,  {\it Int. J. Mod. Phys.} {\bf A 4}, 5047 (1989).
%
\bibitem{Nottale1992} L. Nottale,  {\it Int. J. Mod. Phys.} {\bf A 7}, 4899 (1992).
 %
 \bibitem{Nottale1993} L. Nottale,  {\it Fractal Space-Time and Microphysics: Towards a Theory of Scale Relativity}, World Scientific, Singapore (1993).  %LIWOS
%
\bibitem{Nottale2011}  L. Nottale,  {\it Scale Relativity and Fractal Space-Time: a New Approach to Unifying Relativity and Quantum Mechanics}, Imperial College Press, London (2011).  %ICP
%
 \bibitem{Nottale2007} L. Nottale \& M.N. C\'el\'erier, {\it J. Phys. A: Math. Theor.}, {\bf 40}, 14471 (2007). %Fondation axioms MQ
 %
\bibitem{Nottale2006} L. Nottale L., M.N. C\'el\'erier, T. Lehner, {\it J. Math. Phys.} {\bf 47}, 032303 (2006). % Gauge Field Theory
%
\bibitem{Nottale1996} L. Nottale, {\it Astron. Astrophys. Lett.} {\bf 315}, L9 (1996).
%
\bibitem{Nottale1997} L. Nottale, 1997, {\it Astron. Astrophys.} {\bf 327}, 867 (1997).
%
\bibitem{Auffray2008} C. Auffray and L. Nottale, {\it Progress in Biophysics and Molecular Biology}, {\bf 97}, 79.
%
\bibitem{Nottale2008} L. Nottale and C. Auffray, {\it Progress in Biophysics and Molecular Biology}, {\bf 97}, 115.
%
\bibitem{Turner2015} P. Turner and L. Nottale,  {\it Physica C}, {\bf 515}, 15 (2015). %HTSC 
%
\bibitem{Nottale2018} L. Nottale and T. Lehner, {\it Physics of Fluids} {\bf 31}, 105109 (2019), arXiv:1807.11902.
%
\bibitem{Karplus1950} R. Karplus and N.M. Kroll, {\it Phys. Rev.} { \bf 77}, 536 (1950).
%
\bibitem{Suura1957} H. Suura and E.H. Wichmann, {\it Phys. Rev.} {\bf 105}, 1930 (1957).
%
\bibitem{Petermann1957} A. Petermann, {\it Phys. Rev.} {\bf 105}, 1931 (1957).
%
\bibitem{Elend1966} H.H. Elend, {\it Phys. Lett.} {\bf 20}, 682 (1966).
%
\bibitem{Passera2007} M. Passera, {\it Phys. Rev.} {\bf D 75}, 013002 (2007) .
%
\bibitem{Li1993} G. Li, R. Mendel and M.A. Samuel, {\it Phys. Rev.} {\bf D 47}, 1723 (1993) .
%
\bibitem{Dyson1949} F.J. Dyson, {\it Phys. Rev.} {\bf 75}, 1736 (1949) .
%
\bibitem{Lautrup1974} B.E. Lautrup and E. de Rafael, {\it Nuclear Phys.} {\bf B 70}, 317 (1974).
%
\bibitem{Borsanyi2021} S. Borsanyi, Z. Fodor, J.N. Guenther et al., {\it Nature} (2021). https://doi.org/10.1038/s41586-021-03418-1
%
\bibitem{Grange2015} J. Grange et al. [Muon g-2 Collaboration], arXiv:1501.06858 [physics.ins-det].
 %
 \bibitem{Keshavarzi2020} A. Keshavarzi, D. Nomura, T. Teubner, {\it Phys. Rev.}  {\bf D 101}, 014029 (2020).
 %
 
\end{thebibliography}
\end{document}